\documentclass[fleqn,usenatbib]{mnras}

\usepackage[T1]{fontenc}

\DeclareRobustCommand{\VAN}[3]{#2}
\let\VANthebibliography\thebibliography
\def\thebibliography{\DeclareRobustCommand{\VAN}[3]{##3}\VANthebibliography}

\usepackage{graphicx}	
\usepackage{amsmath}	
\usepackage{amssymb}	
\usepackage{multicol}        
\usepackage{bm}		
\usepackage{pdflscape}	
\usepackage{hyperref}	

\newcommand{\de}{\mathrm{d}}

\usepackage{newtxtext,newtxmath}

\title[Constraints on $\Lambda$CDM from EVS]{AMICO galaxy clusters in KiDS-DR3: constraints on $\Lambda$CDM from extreme value statistics}

\author[V. Busillo et al.]{V. Busillo,$^{1,2,3}$\thanks{E-mail: valerio.busillo@inaf.it}
G. Covone,$^{1,2,3}$
M. Sereno,$^{4,5}$
L. Ingoglia,$^{4}$
M. Radovich,$^{6}$
S. Bardelli,$^{5}$
G. Castignani,$^{4,5}$
\newauthor
C. Giocoli,$^{4,5,7}$
G. F. Lesci,$^{4,5}$
F. Marulli,$^{4,5,7}$
M. Maturi,$^{8,9}$
L. Moscardini,$^{4,5,7}$
E. Puddu,$^{2}$
M. Roncarelli,$^{5}$
\\
$^{1}$Dipartimento di Fisica “E. Pancini”, Universit\`{a} di Napoli Federico II, C.U. di Monte Sant’Angelo, via Cintia, I-80126 Napoli, Italy\\
$^{2}$INAF, Osservatorio Astronomico di Capodimonte, Salita Moiariello 16, I-80131, Napoli, Italy\\
$^{3}$INFN, Sez. di Napoli, Compl. Univ. Monte S. Angelo, Via Cinthia, I-80126 Napoli, Italy\\
$^{4}$Dipartimento di Fisica e Astronomia ``Augusto Righi'' - Alma Mater Studiorum Universit\`{a} di Bologna, via Piero Gobetti 93/2, I-40129 Bologna, Italy\\
$^{5}$INAF, Osservatorio di Astrofisica e Scienza dello Spazio di Bologna, via Piero Gobetti 93/3, I-40129 Bologna, Italy\\
$^{6}$INAF, Osservatorio Astronomico di Padova, vicolo dell'Osservatorio 5, I-35122 Padova, Italy\\
$^{7}$INFN, Sezione di Bologna, viale Berti Pichat 6/2, I-40127 Bologna, Italy\\
$^{8}$Zentrum f\"ur Astronomie, Universit\"at Heidelberg, Philosophenweg 12, D-69120 Heidelberg, Germany\\
$^{9}$ITP, Universit\"at Heidelberg, Philosophenweg 16, D-69120 Heidelberg, Germany\\
}

\date{Accepted 2023 July 18. Received 2023 June 19; in original form 2023 March 7}

\pubyear{2023}

\makeatother

\begin{document}
\label{firstpage} \pagerange{\pageref{firstpage}--\pageref{lastpage}}
\maketitle


\begin{abstract}
We constrain the $\Lambda$CDM cosmological parameter $\sigma_{8}$ by applying the extreme value statistics for galaxy cluster mass on the AMICO KiDS-DR3 catalog. We sample the posterior distribution of the parameters by considering the likelihood of observing the largest cluster mass value in a sample of $N_{\textrm{obs}} = 3644$ clusters with intrinsic richness $\lambda^{*} > 20$ in the redshift range $z\in[0.10, 0.60]$. We obtain $\sigma_{8}=0.90_{-0.18}^{+0.20}$, consistent within $1\sigma$ with the measurements obtained by the Planck collaboration and with previous results from cluster cosmology exploiting AMICO KiDS-DR3. The constraints could improve by applying this method to forthcoming missions, such as \textit{Euclid} and LSST, which are expected to deliver thousands of distant and massive clusters.
\end{abstract}



\section{Introduction}
The extreme value statistics (EVS, \citealt{Gumbel1958}) seeks to determine how likely the highest valued observations of a random variable is. In the last decades, this theory has seen various applications in astrophysics and cosmology.
For instance, it has been used to analyze the luminosity distribution of the most massive galaxies in a galaxy cluster \citep{Bhavsar1985}, 
the convective penetration in radiative-convective boundaries of pre-main sequence stars \citep{Pratt2017}, and the most massive halos in the Universe \citep{Sheth2011,Waizmann2011,Waizmann2012,HolzPerlmutter2012}.
In particular, it has been used to predict the probability distribution for the most massive galaxy cluster in a given region of the Universe, with results that are highly consistent with predictions from large, well resolved cosmological $N$-body simulations \citep{Davis2011}.

In cosmology, this approach has been used to verify whether the observation of rare astrophysical objects is compatible with the predictions from a cosmological model (for example, the $\Lambda$-cold dark matter standard cosmological model, $\Lambda$CDM). \cite{Harrison2011} applied the extreme value statistics to galaxy clusters to check if the most massive galaxy clusters were in agreement with the standard cosmological model or with alternative models that predict an enhanced structure formation, with all the clusters considered showing concordance with a $\Lambda$CDM cosmology. \cite{Kim2021} used the same approach as a rarity test to check whether the physical properties of the galaxy cluster ACT-CL J0102–4915 (also known as `El Gordo') were compatible with the $\Lambda$CDM paradigm, finding its mass compatible with $\Lambda$CDM predictions within $2\sigma$. 

More recently, \cite{Lovell2022} applied the extreme value statistics to Hubble Space Telescope (HST) and James Webb Space Telescope (JWST) observations of high-redshift galaxies, to verify whether these objects were in tension with the standard cosmological model. They found significant tension with $\Lambda$CDM predictions for some $z\gtrsim 10$ galaxies taken from the recent JWST high-redshift candidates.

An approach using the extreme value statistics with respect to cosmological parameters was also studied in \cite{Reischke2016}, where the asymptotic limit of the extreme value statistics distribution for a large number of observations, the generalized extreme value distribution (GEV), has been applied to mock weak lensing shear peak counts from an \textit{Euclid}-like survey to find confidence regions for the amplitude of the linear matter density fluctuations $\sigma_{8}$, the matter density parameter, $\Omega_{\textrm{m}}$, and the dark energy equation of state parameter, $w_{0}$.

The inferred values of $S_{8}\equiv \sigma_{8} \sqrt{\Omega_{\textrm{m}}/0.3}$ indicate a discrepancy, as demonstrated in \cite{Douspis2019} and \cite{Corasaniti2021}, between late-time studies, such as cosmic shear measurements \citep{Asgari2021, Hikage2019} and early-epoch investigations, for example those derived from primary CMB analysis \citep{Planck2013, Planck2018_I}. This tension is significant at a $2$-$4\sigma$ level (refer to \citealt{Abdalla2022} for a review). Consequently, it is crucial to examine how findings based primarily on cluster mass measurements compare to this tension.

Here, we apply EVS to the highest value of galaxy cluster mass in a survey, to constrain the value of $\sigma_{8}$. We consider the cluster catalog presented in \cite{Maturi2019}, which was built by using the Adaptive Matched Identifier of Clustered Objects (AMICO) algorithm \citep{Bellagamba2018} on the Third Data Release of the Kilo Degree Survey (KiDS-DR3, \citealt{DeJong2017}).

The paper is organized as follows. In Section \ref{sec:The_data} we briefly present the AMICO KiDS-DR3 cluster catalog. In Section \ref{sec:EVS_for_cluster_masses}, we construct the probability distribution for the mass of the most massive cluster expected in the catalog. In Section \ref{sec:Statistical_model}, we describe the likelihood on which we perform the Bayesian inference of the relevant cosmological parameters and compare the observations from the AMICO KiDS-DR3 catalog to the theoretical predictions. The results of the analysis are presented in Section \ref{sec: Results}, and are discussed in Section \ref{sec:Discussion}. Section \ref{sec: Systematic_errors} is dedicated to the analysis of systematics, such as the uncertainty due to the mass function model and the uncertainty related to the mass ranking. Lastly, our conclusions are presented in Section \ref{sec: Conclusions}.

All the cluster masses considered in this work are defined using $M_{200c}$, i.e. the mass inside a sphere that encloses an average mass density equal to 200 times the critical density of the Universe at the cluster redshift\footnote{$\rho_c (z) = 3H(z)^2/8 \pi G$}. We also define $\log_{10}[M_{200c}\,(10^{14}\,h^{-1}\textrm{M}_{\odot})] = \mu$, for brevity. In this work, we assume a flat $\Lambda$CDM cosmology with only two free parameters, $\sigma_{8}$ and $\Omega_{\textrm{m}}$. All the other parameters, for example the dark energy density parameter, $\Omega_{\Lambda}$, the dark energy equation of state parameter, $w$, the Hubble constant, $H_0$, and the spectral index, $n_{\textrm{s}}$, are fixed to the values estimated in \cite{Planck18}, Table 2 (TT, TE, and EE+lowE). All the results shown in this work are given in terms of 16th, 50th (median) and 84th percentiles of the respective distributions.

\section{The AMICO KiDS-DR3 cluster catalog}\label{sec:The_data}

The clusters used in this work were detected in the third data release (DR3) of KiDS \citep{DeJong2013,Kuijken2015, DeJong2017}. KiDS is an European Southern Observatory (ESO) wide-field imaging survey made with the OmegaCAM camera \citep{Kuijken2011} mounted on the Very Large Telescope (VLT) Survey Telescope (VST, \citealt{CapaccioliSchipani2011}), designed to observe in the $(u,g,r,i)$ bands an area of 1350 $\textrm{deg}^{2}$, down to the limiting magnitudes of 24.3, 25.1, 24.9, and 23.8 for each of the bands, respectively. The DR3 covers an area of 438 $\textrm{deg}^{2}$ (377 $\textrm{deg}^{2}$ after removing areas affected by artifacts such as reflection halos around bright stars and satellite tracks, see \citealt{Jong2015, Kuijken2015}).

\cite{Maturi2019} presented a catalog of clusters found in KiDS-DR3 with AMICO, a code designed for the detection of galaxy clusters in photometric surveys, first tested on mock photometric galaxy catalogs derived in \cite{Bellagamba2018}. The detection algorithm is based on the Optimal Filtering technique \citep{Maturi2005,Bellagamba2011,Radovich2017}. As detailed in \cite{Maturi2019}, AMICO models the data as the sum of a signal term and a noise term, and uses an iterative method to evaluate the signal amplitude by filtering the raw data with a filter evaluated on a 3D grid $(\boldsymbol{\theta}_{c},z_{c})$, where $\boldsymbol{\theta}_{c}$ is the sky coordinate vector, given in terms of right ascension and declination, and $z_{c}$ is the redshift \citep{Bellagamba2018}. As grid resolution, \cite{Maturi2019} considered $0.3\;\textrm{arcmin}$ for each of the $\boldsymbol{\theta}_{c}$ components and $0.01$ for the redshift.

A cluster is localized by considering the location in the `sky coordinates-redshift' space $(\boldsymbol{\theta}_{c},z_{c})$ which maximizes the amplitude $A(\boldsymbol{\theta}_{c},z_{c})$. The model used to describe the signal term is the product of a Schechter luminosity function \citep{Schechter1976} by a Navarro-Frenk-White radial density profile \citep{Navarro1997}. Once a detection is defined, AMICO also assigns to each galaxy a membership probability to belong to the $j$-th cluster, based on the value of the amplitude signal $A$ and on the observed properties (sky position, photometric redshift distribution and magnitude) of the galaxy.

It should be noted that, as described in \cite{Bellagamba2018}, AMICO can be considered a 3D detection algorithm, since it considers the photometric redshift distribution of each galaxy when computing the amplitude and S/N of each detection. In particular, a cleaning procedure was applied, which was specifically designed to minimize blending and projection effects.

Among the different mass proxies provided by AMICO, in this work we used the intrinsic richness $\lambda^{*}$ as the mass proxy, defined as:

\begin{equation}
\lambda_{j}^{*}=\sum_{i=1}^{N_{\textrm{gal}}}P_{i}(j)\quad\textrm{with }\begin{cases}
m_{i}<m^{*}(z_{j})+1.5\\
R_{i}(j)<R_{\textrm{max}}(z_{j})
\end{cases} \, ,
\end{equation}
\\
where $P_{i}(j)$ is the membership probability for the $j$-th cluster located  at redshift $z_{j}$, $m_{i}$ is the magnitude of the galaxy, $m^{*}$ the Schechter function reference magnitude, $R_{i}$ is the distance of the galaxy
from the center of the cluster and $R_{\textrm{max}}$ is the radius of a sphere enclosing a mass $M_{200c}=10^{14}h^{-1}\,\textrm{M}_{\odot}$. This choice is made because the intrinsic richness is a reliable halo mass proxy,  which is fundamental for cosmological studies using clusters.

The mass-richness relation used to convert the mass proxy is:

\begin{equation}
\mu_{\chi}(z;\lambda^{*})=\alpha+\beta\log_{10}\left(\frac{\lambda^{*}}{\lambda_{\textrm{piv}}^{*}}\right)+\gamma\log_{10}\left(\frac{E(z)}{E(z_{\textrm{piv}})}\right)\, , \label{eq:mass_richness_scatter_relation}
\end{equation}
\\
where $\alpha=0.004\pm0.038$, $\beta=1.71\pm0.08$, $\gamma=-1.33\pm0.64$, $\lambda_{\textrm{piv}}^{*}=30$ and $z_{\textrm{piv}}=0.35$. The scatter of this relation is given by $\log_{10}(\sigma_{\mu|\lambda^{*}}) = -1.13\pm 0.53$ \citep{Sereno2020}. The relation was calibrated with weak lensing mass estimates in \cite{Bellagamba2019}, which we refer to for details.

The AMICO KiDS-DR3 catalog \citep{Maturi2019} contains 7988 clusters, down to $S/N=3.5$, in the redshift range $z\in[0.10,0.80]$. For the analysis, we used the cluster redshifts corrected for the bias discussed in \cite{Maturi2019},  and we ranked the cluster masses $M_{200c}$ based on the respective intrinsic richness $\lambda^{*}$.

\begin{figure}
\centering
\includegraphics[width=\columnwidth]{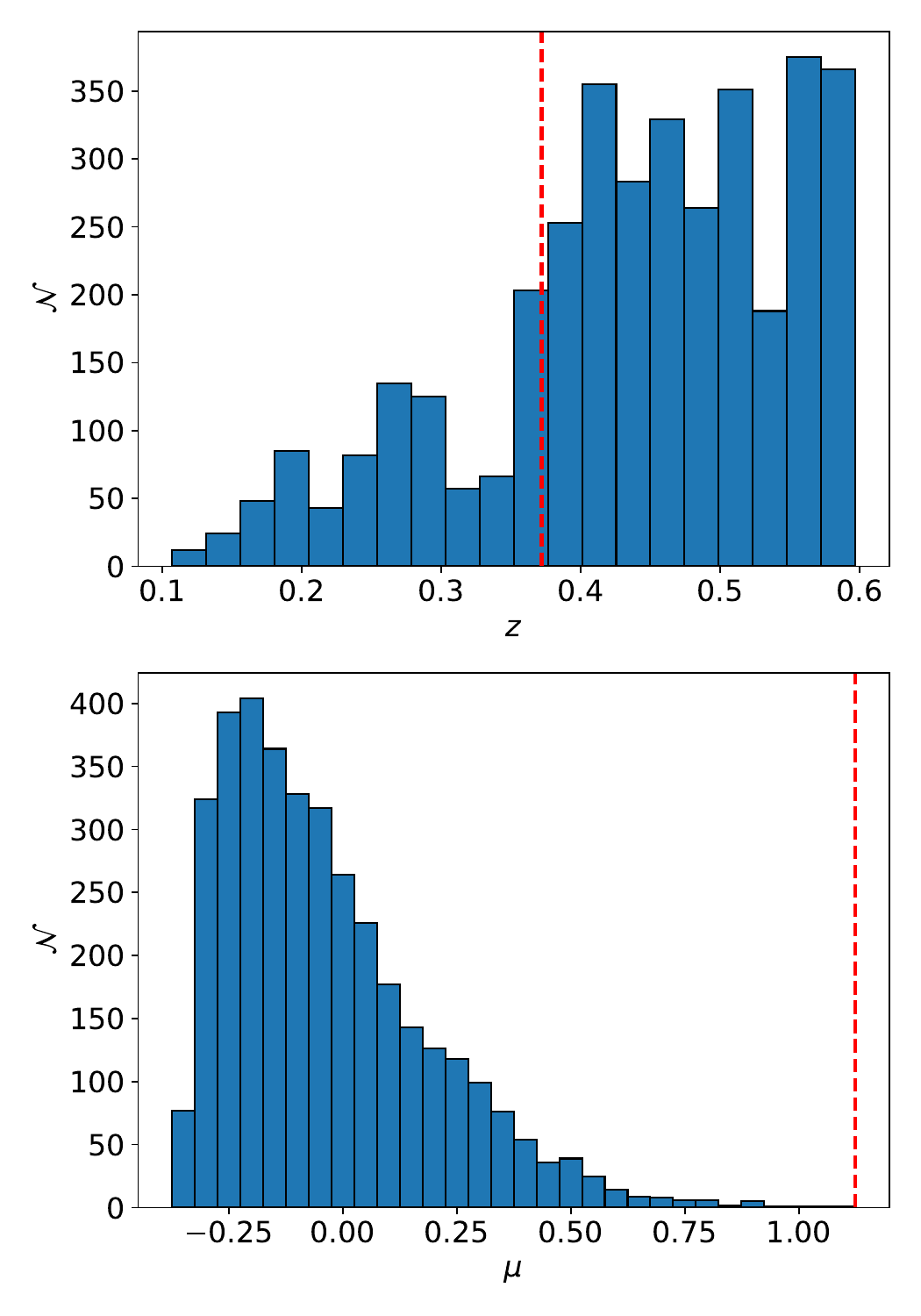}\caption{\label{fig:Redshift_mass_distribution}Redshift and mass distributions of the clusters in the AMICO KiDS-DR3 catalog having $\lambda^{*} > 20$ and $z$ in the redshift range $[0.10, 0.60]$. \textit{Top panel}: redshift distribution. \textit{Bottom panel}: mass distribution. The dashed vertical red lines represent the redshift and mass associated with the most massive cluster of the catalog.}
\end{figure}

To be consistent with the scaling relation parameter estimates obtained
in \cite{Bellagamba2019}, we restricted ourselves to the redshift
interval $z\in[0.10, 0.60]$ and filtered out all the clusters with
intrinsic richness $\lambda^{*}\leq 20$, leaving $N_{\textrm{obs}}=3644$ clusters. This ensures a purity of the
sample of more than 98 per cent over the whole interval, and a completeness
of $\sim 84$ per cent in the redshift range $z\in[0.10, 0.30]$, $\sim 79$ per cent
in the range $z\in[0.30, 0.45]$ and $\sim 58$ per cent in the range $z\in[0.45,0.60]$
(see \citealt{Maturi2019}, fig. 12). Fig. \ref{fig:Redshift_mass_distribution} shows the redshift and
mass distributions of the galaxy clusters in our sample.

\begin{figure}
\centering
\includegraphics[width=\columnwidth]{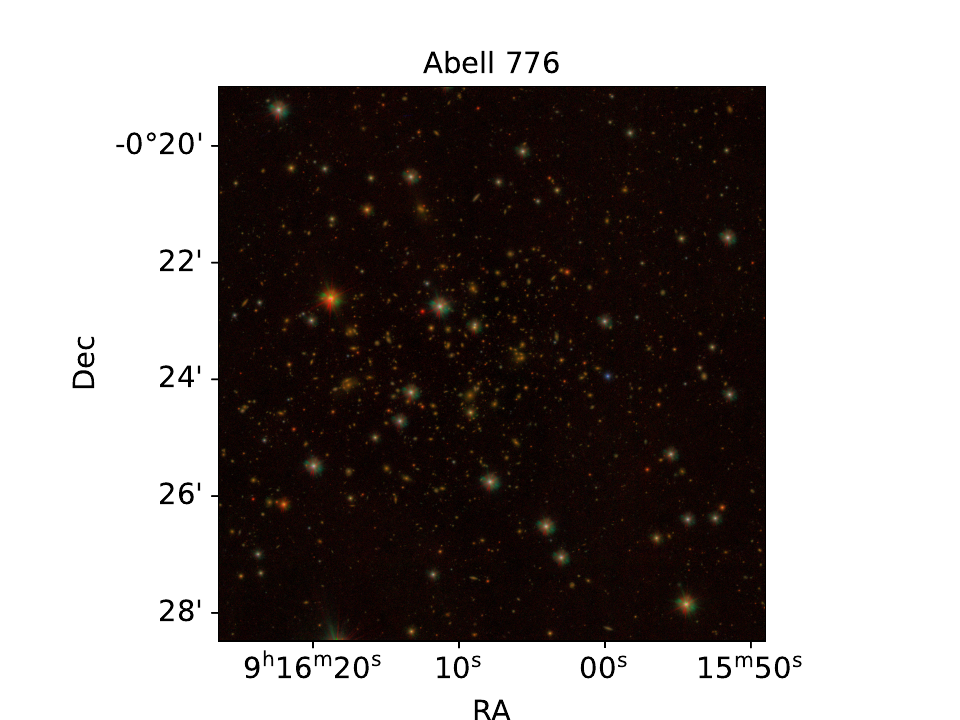}
\caption{\label{fig:Abell776}Colour composite $(g,r,i)$ image relative to J091606.48-002328, also known as Abell 776. The stamp shows a $\approx 6' \times 6'$ region centered on the location identified by the AMICO algorithm.}
\end{figure}

The most massive cluster in AMICO KiDS-DR3 is AK3 J091606.48-002328, also known as Abell 776. This cluster is located at redshift $z=0.37$, and has an intrinsic richness of $\lambda^{*}=137\pm24$, with a corresponding mass of $M_{200c}=(13.3\pm4.9)\times 10^{14}h^{-1}\,\textrm{M}_{\odot}$. The corresponding log-mass is $\mu = 1.12\pm 0.16$. This cluster is also present in \cite{Hamana2020} and \cite{Medezinski2018}, with masses reported in Table \ref{table: Abell776_masses}. Fig. \ref{fig:Abell776} shows an image of Abell 776.

As detailed in \cite{Sereno2015a}, weak lensing mass estimates depend on the cosmological parameters via the equation:
\begin{align}
M_{\Delta}^{\textrm{WL}}\propto &\; D_{\textrm{d}}^{-3\delta_{\gamma}/(2-\delta_{\gamma})}\left(\frac{D_{\textrm{ds}}}{D_{\textrm{s}}}\right)^{-3/(2-\delta_{\gamma})}\nonumber \\
 & \times\; \left[H(z)\right]^{-(1+\delta_{\gamma})/(1-\delta_{\gamma}/2)}\,,\label{eq:observed_mass_cosmology_dependence}
\end{align}
\\
where $z$ is the redshift of the cluster, $\delta_{\gamma}$ is a parameter related to the slope of the angular mass profile, $D_{\textrm{d}}$ is the observer-lens angular-diameter distance, $D_{\textrm{s}}$ is the observer-source angular-diameter distance and $D_{\textrm{ds}}$ is the lens-source angular-diameter distance.
Therefore, when changing the underlying cosmology, one should correct the observed mass value via the formula:

\begin{equation}
M_{\Delta,\textrm{new}}=\frac{\alpha_{\textrm{new}}}{\alpha_{\textrm{old}}}M_{\Delta,\textrm{old}}\,,
\end{equation}
\\
where $\alpha_{\textrm{new}}$ is the factor defined in equation \eqref{eq:observed_mass_cosmology_dependence}, evaluated with
the new choice of cosmological parameters, while $\alpha_{\textrm{old}}$ is the same factor, but evaluated assuming the reference cosmology.

\begin{table}
\centering
\renewcommand*{\arraystretch}{1.50}

\caption{Mass estimates of Abell 776, along with the measurement method used to determine them.}

\begin{tabular}{ccc}
\hline
\hline
$M_{200c}$ & Method & Reference\\
$[10^{14}h^{-1}\,\textrm{M}_{\odot}]$ & &\\
\hline
$13.3\pm 4.9$ & Richness & This work\\
$5.61_{-1.68}^{+1.96}$ & Weak Lensing & \cite{Hamana2020}\\
$8.10_{-2.17}^{+3.26}$ & Weak Lensing & \cite{Medezinski2018}\\
$4.33_{-0.56}^{+0.58}$ & Sunyaev-Zel'dovich & \cite{Planck2016}\\
\hline 
\end{tabular}

\label{table: Abell776_masses}
\end{table}

\section{EVS for cluster masses}\label{sec:EVS_for_cluster_masses}

The probability distribution of the largest cluster mass, $M_{\textrm{max}}$, in a sample $\{M_{i}\}$ of $N$ clusters, i.e. the maximum mass $M_{\textrm{max}} = \textrm{max}(M_{i})$, was discussed e.g. in \cite{Harrison2011} and \cite{Waizmann2013}.

If all the $M_{i}$ values are drawn from the same probability
distribution, and the measurements are independent, the
probability density function $\phi:\de P(M_{\textrm{max}}=M|N)=\phi(M)\,\de M$
associated to the highest order statistics in mass is given by (\citealt{Waizmann2013}, appendix A1):

\begin{equation}
\phi(M_{\textrm{max}}=M;N)=N\,f(M)\,[F(M)]^{N-1}\label{eq:EVS_phi_distribution} \, ,
\end{equation}
\\
while the associated cumulative distribution function is:

\begin{equation}
\Phi(M_{\textrm{max}}\leq M;N)= \int_{0}^{M}\phi(m)\,dm=[F(M)]^{N}.
\end{equation}

The distributions $f(M)$ and $F(M)$ are the probability density
 function and the cumulative distribution function of a galaxy cluster of mass $M$ in a region of the sky, respectively. These distributions can be written in terms of the halo mass function, $\de n/\de M$, which
describes the number density of clusters expected to have a mass between
$M$ and $M+\de M$, and the comoving volume per unit of redshift, $\de V/\de z$.

Following \cite{Waizmann2013}, these distributions can be written
for a fixed redshift range $[z_{\textrm{min}},\,z_{\textrm{max}}]$ as:

\begin{align}
f(M) & =\frac{f_{\textrm{sky}}}{N_{\textrm{tot}}}\int_{z_{\textrm{min}}}^{z_{\textrm{max}}}\chi(M,z;\lambda_{\textrm{th}}^{*})\,\frac{\de n}{\de M}(M,z)\,\frac{\de V}{\de z}(z)\,\de z,\label{eq:EVS_f_distribution}\\
F(M) & =\int_{M_{\textrm{min}}}^{M}f(m)\,\de m,\label{eq:EVS_F_distribution}
\end{align}
\\
where $f_{\textrm{sky}}$ is the fraction of the sky covered by the survey, and:

\begin{equation}
N_{\textrm{tot}}=f_{\textrm{sky}}\int_{M_{\textrm{min}}}^{M_{\textrm{max}}}\int_{z_{\textrm{min}}}^{z_{\textrm{max}}}\chi(M,z;\lambda_{\textrm{th}}^{*})\,\frac{\de n}{\de M}(M,z)\,\frac{\de V}{\de z}(z)\,\de z\,\de M\,.\label{eq:EVS_Ntot_distribution}
\end{equation}

The function $\chi$ that appears in equations \eqref{eq:EVS_f_distribution}, \eqref{eq:EVS_F_distribution} and \eqref{eq:EVS_Ntot_distribution} is the selection function, which accounts for the fact that it is not possible to detect all galaxy clusters theoretically observable in a certain Universe volume, due to limited depth or accuracy of the detection algorithm. The parameter $\lambda_{\textrm{th}}^{*}$ is the threshold in cluster richness, which we fixed to 20 (see Section \ref{sec:The_data}).

The selection function $\chi$ can be approximated as \citep{Sereno2015}:

\begin{equation}
\chi(\mu,z;\lambda_{\textrm{th}}^{*})=\frac{1}{2}\textrm{erfc}\left(\frac{\mu_{\chi}(z;\lambda_{\textrm{th}}^{*})-\mu}{\sqrt{2}\sigma_{\chi}}\right) \, ,\label{eq:EVS_selection_function}
\end{equation}
\\
the scale $\sigma_{\chi}$ can be written in terms of uncertainty on mass estimates, obtained by adding in quadrature the intrinsic scatter of the mass-richness relation to the uncertainty associated to the scaling parameters in equation \eqref{eq:mass_richness_scatter_relation}, and the uncertainty associated to the richness, $\delta\lambda^{*}/\lambda^{*}\sim 18$ per cent \citep{Maturi2019}.

As a reference, we consider the halo mass function from \cite{Despali2016}. In Section \ref{subsec: HMF_systematics} we will discuss the possible systematics introduced by this choice.
We used a single redshift interval for evaluating equations (\ref{eq:EVS_f_distribution}-\ref{eq:EVS_Ntot_distribution}), fixing $z_{\textrm{min}}=0.10$ and $z_{\textrm{max}}=0.60$.

\section{Statistical model} \label{sec:Statistical_model}
In the following, we consider a flat $\Lambda$CDM cosmological model with only two free parameters, $\sigma_{8}$ and $\Omega_{\textrm{m}}$. The observed quantities are two: the observed log-mass of Abell 776, $\mu_{\textrm{max, obs}}$, and the observed cluster count, $N_{\textrm{obs}}$. The model parameters are four: the $\Lambda$CDM cosmological parameters $\sigma_{8}$ and $\Omega_{\textrm{m}}$, the true log-mass of Abell 776, $\mu_{\textrm{max, true}}$, and the true cluster count, $N_{\textrm{true}}$.

The likelihood distribution is the probability of the largest mass estimate in the survey and of the observed number cluster count for a given set of cosmological parameters and the true values of the mass and of the cluster count.

We used the following functional form for the likelihood:

\begin{align}
L=\mathcal{G}(\mu_{\textrm{max, obs}}|\,\mu_{\textrm{max, true}},\,\sigma_{\mu})\cdot & \mathcal{P}(N_{\textrm{obs}}|\,N_{\textrm{true}})\nonumber \\
\times\; \phi(\mu_{\textrm{max, true}}|\, & N_{\textrm{true}},\,\sigma_{8},\,\Omega_{\textrm{m}})\;, \label{eq: EVS_likelihood}
\end{align}
\\
which is the product of three distributions:

\begin{itemize}
\item A Gaussian distribution, $\mathcal{G}$, centered on the expected log-mass value $\mu_{\textrm{max, true}}$ and evaluated for $\mu=\mu_{\textrm{max, obs}}$, with a standard deviation equal to the statistical uncertainty on the mass;

\item A Poisson distribution, $\mathcal{P}$, with expected value equal to $N_{\textrm{true}}$ and evaluated for $N = N_{\textrm{obs}}$. This term takes into account the probability of observing a cluster count equal to $N_{\textrm{obs}}$;

\item The extreme value statistics distribution, $\phi$, evaluated for $\mu = \mu_{\textrm{max, true}}$, which gives the probability of observing a given log-mass for the most massive cluster, in a region of the sky where $N_{\textrm{true}}$ clusters are located, having a log-mass equal to $\mu_{\textrm{max, true}}$.
\end{itemize}

We want to remark that our approach is complementary to the one commonly used in cosmological inference from cluster abundance, where the number count is written in terms of the halo mass function and the cosmological parameters (see e.g. equation \ref{eq:EVS_Ntot_distribution}), in that we instead treat $N_{\textrm{true}}$ as a model parameter to be marginalized over.

Finally, to estimate the full posterior distribution, we considered uniform priors, $\mathcal{U}$, for all the parameters, with bounds reported in Table \ref{table: Results_fixed_ns}.

\subsection{Compatibility of AMICO KiDS-DR3 clusters with \texorpdfstring{$\Lambda$CDM}{L-CDM} predictions}
We verified whether the mass measurements of AMICO KiDS-DR3 clusters are compatible with the theoretical predictions from a $\Lambda$CDM cosmology having all the cosmological parameters fixed to the \cite{Planck18} results.

We evaluated equations (\ref{eq:EVS_f_distribution}-\ref{eq:EVS_Ntot_distribution}) in redshift bins of width $\Delta z = 0.1$, obtaining confidence intervals for the most massive galaxy cluster mass expected by the cosmological model, given by the equation:

\begin{equation}
\Phi(M,z) = \textrm{cost.}\, ,
\end{equation}
\\
where the constant is equal to a certain quantile.

\begin{figure}
\centering
\includegraphics[width=\columnwidth]{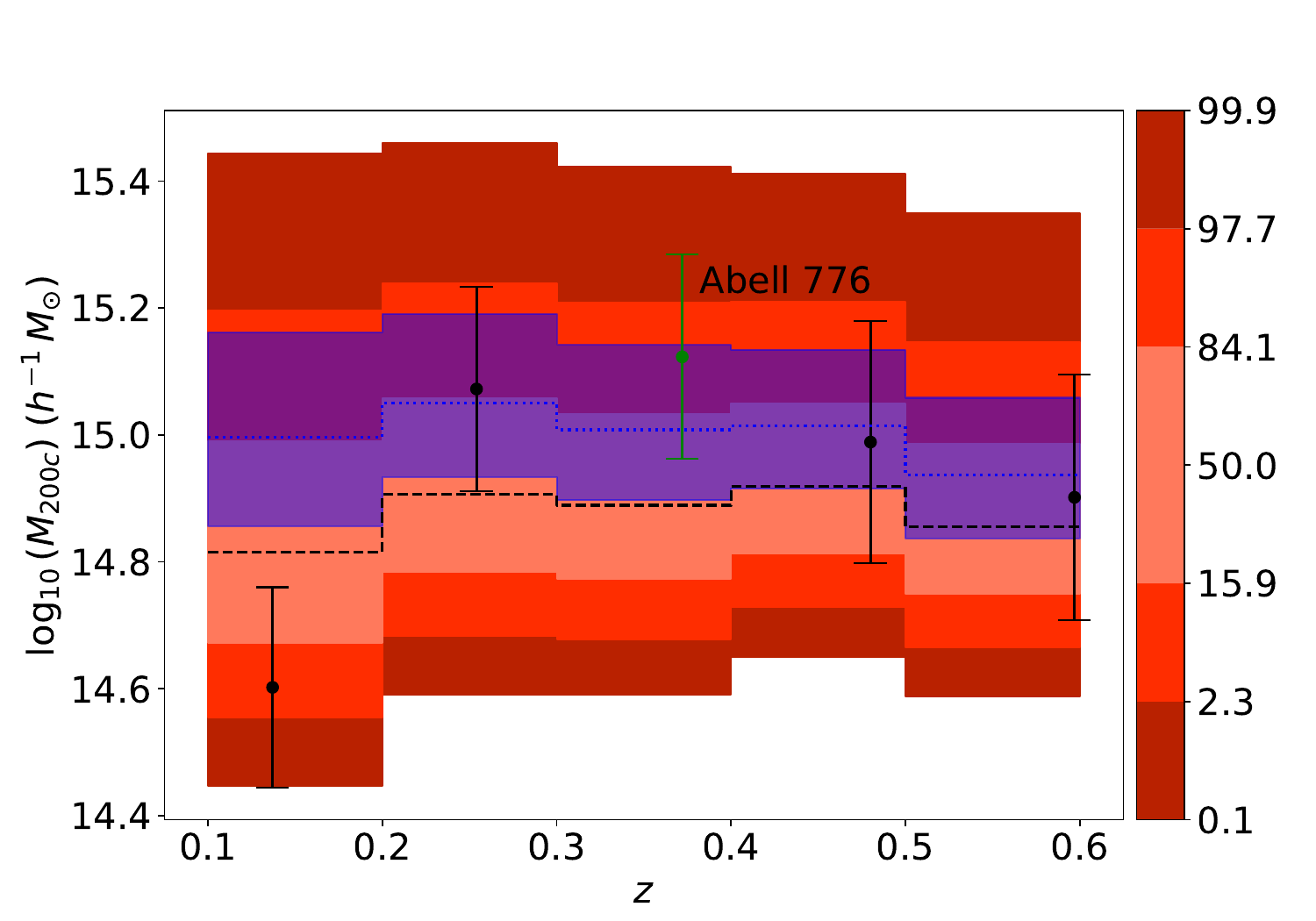}
\caption{\label{fig:EVS_contours_Abell776}Extreme value statistics predictions in mass-redshift space for a $\Lambda$CDM cosmology having cosmological parameters fixed to the \protect\cite{Planck18} results (red contours for each redshift bin $\Delta z = 0.1$), compared with the observed largest cluster mass values from the AMICO KiDS-DR3 catalog (black points). The green point is Abell 776's mass measurement. We also show the $[\textrm{16th, 84th}]$ confidence interval predictions for a cosmology based on the maximum likelihood parameters associated to equation \eqref{eq: EVS_likelihood}.}
\end{figure}

In Fig. \ref{fig:EVS_contours_Abell776}, we show with red contours the predictions for the largest cluster mass value for each redshift bin, and we compare them with the observed most massive cluster mass in each bin (black points). The observed Abell 776 mass (green point) is compatible with a $\Lambda$CDM cosmology having the cosmological parameters fixed to the results from \cite{Planck18} within $1\sigma$.

\section{Results}\label{sec: Results}

We constrained the posterior with a Monte Carlo Markov Chain (MCMC) analysis. As sampler, we used \textsc{emcee} \citep{ForemanMackey2013}, with 32 walkers and running a chain of 1000 steps, equal to $\sim 27$ times the autocorrelation time of the chain, starting with initial positions for the walkers extracted from a Gaussian distribution centered around $\sigma_{8}=0.810$ and $\Omega_{\textrm{m}}=0.311$ based on \cite{Planck18} results, $\mu_{\textrm{max, true}}=1.12$ and $N_{\textrm{true}}=3644$ based on the values of $\mu_{\textrm{max, obs}}$ and $N_{\textrm{obs}}$. To minimize any possible influence from the starting position on the final results, we discarded the first 100 steps. We also thinned the sample with a step equal to $0.5$ times the autocorrelation time. To further check for convergence, we verified that the results do not change for shorter chains.

\begin{figure*}
\centering
\includegraphics[width = \textwidth]{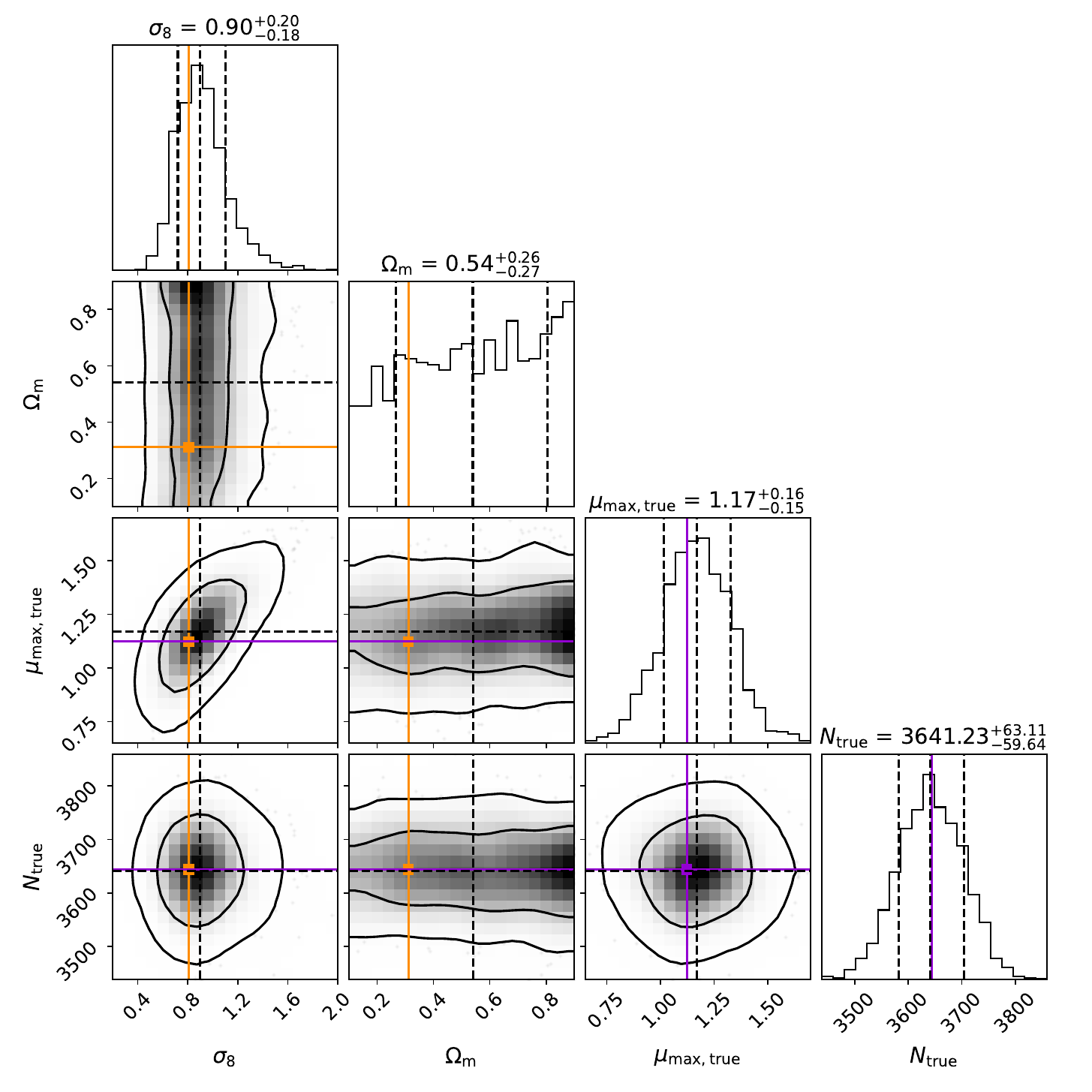}
\caption{\label{fig:corner_plot_fixed_ns}Corner plot showing the 16th, 50th and 84th percentiles of the 1D parameter distributions with the $68.3$ per cent and $95.4$ per cent confidence regions for the corresponding 2D histograms, obtained by sampling the posterior probability distribution. Our results are shown with dashed black lines. The \protect\cite{Planck18} results and the observed values $\mu_{\textrm{max, obs}}$ and $N_{\textrm{obs}}$ are shown in orange and violet, respectively.}
\end{figure*}

We obtained $\sigma_{8}=0.90_{-0.18}^{+0.20}$ and $\Omega_{\textrm{m}}=0.54_{-0.27}^{+0.26}$, as reported in Table \ref{table: Results_fixed_ns}. The posterior is shown in Fig. \ref{fig:corner_plot_fixed_ns}. While the results on $\sigma_{8}$ are encouraging, showing good constraints on the parameter, we are unable to provide meaningful constraints on $\Omega_{\textrm{m}}$.

We also considered the constraints on $S_{8} \equiv \sigma_{8}\sqrt{\Omega_{\textrm{m}}/0.3}$. The resulting distribution is shown in Fig. \ref{fig:S8_distribution}. We found $S_{8} = 1.16_{-0.36}^{+0.40}$, also shown in Table \ref{table: Results_fixed_ns}.
\begin{table}
\centering
\renewcommand*{\arraystretch}{1.50}

\caption{Parameters associated to the likelihood distribution described in Section \ref{sec:Statistical_model}.}

\begin{tabular}{ccc}
\hline
\hline
Parameter & Prior & Posterior\\
\hline
$\sigma_{8}$ & $\mathcal{U}(0, 2)$ &$0.90_{-0.18}^{+0.20}$\\
$\Omega_{\textrm{m}}$ & $\mathcal{U}(0.10,0.90)$ & $0.54_{-0.27}^{+0.26}$\\
$S_{8}$& $-$ & $1.16_{-0.36}^{+0.40}$ \\
$\mu_{\textrm{max,true}}$&$\mathcal{U}(-2,4)$& $1.17_{-0.15}^{+0.16}$\\
$N_{\textrm{true}}$ &$\mathcal{U}(0,10^{4})$& $3641_{-60}^{+63}$\\
\hline 
\end{tabular}

\label{table: Results_fixed_ns}
\end{table}

\begin{figure}
\centering
\includegraphics[width=\columnwidth]{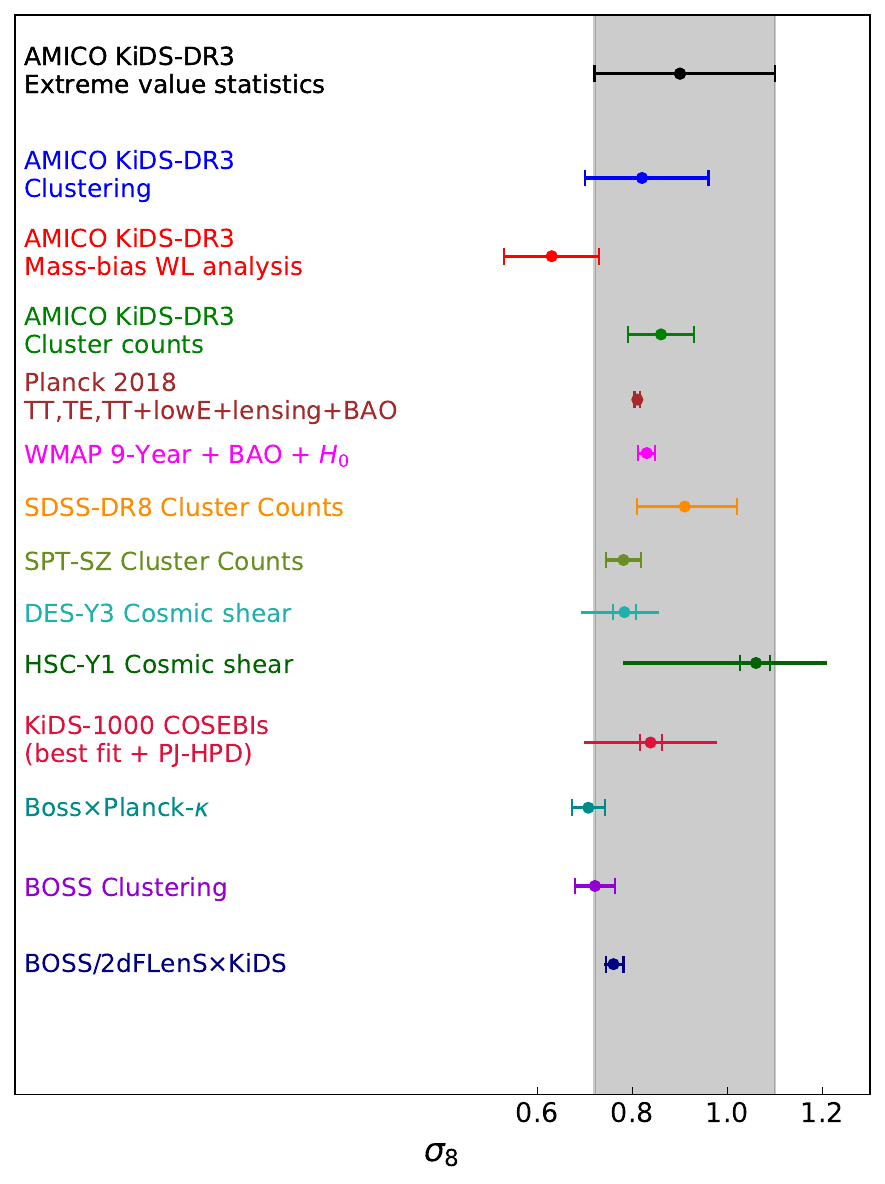}
\caption{\label{fig:sigma8_measurements_comparison}Comparison between $\sigma_{8}$ measurements obtained, from top to bottom, by using the posterior distribution derived from the likelihood of observing the most massive cluster in the AMICO KiDS-DR3 catalog (black dot), from the analysis of the 2PCF on AMICO KiDS-DR3 catalog presented in \protect\cite{Lesci2022_clustering}, from the fitting of the mass-bias relation performed in \protect\cite{Ingoglia2022}, from the cluster counts/weak lensing joint analysis presented in \protect\cite{Lesci2022} and from the results presented in \protect\cite{Planck18}, \protect\cite{Hinshaw2013}, \protect\cite{Costanzi2019}, \protect\cite{Bocquet2019}, \protect\cite{Amon2022} $\And$ \protect\cite{Secco2022}, \protect\cite{Hikage2019}, \protect\cite{Asgari2021}, \protect\cite{Chen2022}, \protect\cite{Ivanov2020} and \protect\cite{Heymans2021}. The interval between the 16th and 84th percentile in our measurement is also shown in grey. The error lines associated to cosmic shear measurements are drawn with respect to the reported marginalized results, while the error caps are obtained by taking the respective constraints on $S_{8}$ and considering $\Omega_{\textrm{m}}$ fixed to the \protect\cite{Planck18} result.}
\end{figure}

\begin{figure}
\centering
\includegraphics[width=\columnwidth]{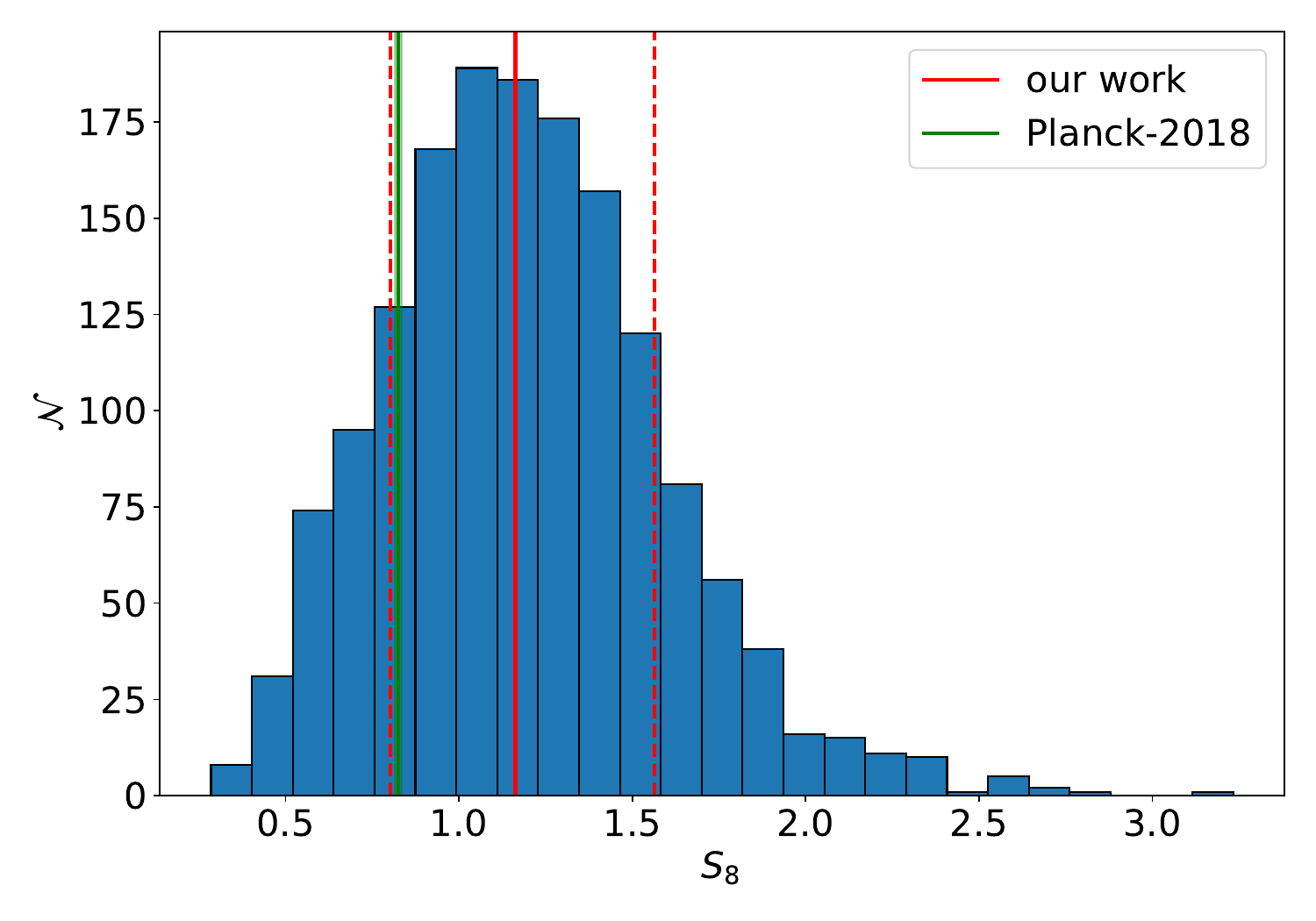}
\caption{\label{fig:S8_distribution}Sampled posterior distribution of $S_{8}$. Dashed vertical red lines show the 16th and 84th percentiles of the distribution around the median (full red line), while the green shadowed region shows the \protect\cite{Planck18} result.}
\end{figure}

\section{Discussion}\label{sec:Discussion}

Given the quite large statistical uncertainty, our results cannot distinguish $\sigma_{8}$ values as estimated either from late or early-time experiments. They are in agreement within $1\sigma$ with Planck results \citep[Table 2, TT, TE, and EE+lowE]{Planck18}. Our constraints are also in agreement within $1\sigma$ with previous results from cluster cosmology exploiting AMICO KiDS-DR3, for example the cluster counts/weak lensing joint analysis \citep{Lesci2022}, the two-point correlation function \citep{Lesci2022_clustering} and the mass-bias relation \citep{Ingoglia2022}. Fig. \ref{fig:sigma8_measurements_comparison} shows a comparison between our measurement and these results.

The relative uncertainty of our $\sigma_{8}$ measurement is 0.21. In comparison, the uncertainties for the \cite{Lesci2022} cluster count and \cite{Lesci2022_clustering} clustering results, both derived from the AMICO KiDS-DR3 catalog, are 0.08 and 0.16, respectively. Although the EVS approach is not yet competitive with cluster count methods, its relative uncertainty is approaching that of clustering.

$\Omega_{\textrm{m}}$ is poorly constrained by the cluster mass EVS. This may be due to the fact that the order statistics with respect to mass is more sensitive to variations of $\sigma_{8}$ rather than of $\Omega_{\textrm{m}}$, as verified also by \cite{Waizmann2013}.

The constraint on $S_{8}$ is also in agreement within $1\sigma$ with \cite{Planck18} results and with previous results from cluster cosmology exploiting AMICO KiDS-DR3, for example the cluster counts/weak lensing joint analysis \citep{Lesci2022} and the two-point correlation function \citep{Lesci2022_clustering}, and within $2\sigma$ from the mass-bias relation analysis \citep{Ingoglia2022}. Differently from other probes based on cluster cosmology, EVS is more sensitive to $\sigma_{8}$ rather than $S_{8}$.

We show in Fig. \ref{fig:EVS_contours_Abell776} the predictions for the $[\textrm{16th, 84th}]$ confidence interval associated to the most massive cluster mass, evaluated by considering the maximum likelihood cosmological parameters obtained from equation \eqref{eq: EVS_likelihood}. The higher $\sigma_{8}$ and $\Omega_{\textrm{m}}$ values with respect to the \cite{Planck18} results have the effect of shifting the predictions towards higher mass values.

The results reported in Section \ref{sec: Results}  have been obtained based on a relatively small sample of clusters. Considering that the extreme value statistics distribution has a variance which decreases with an increasing number of sample objects, we expect an improvement on the estimates of the cosmological parameters by using a larger catalog.

To illustrate the effects of increasing the number of observed objects on the cosmological parameters' uncertainty, we repeated the analysis after modifying the likelihood, multiplying the number of observed clusters in the catalog and the covered area of the sky by a factor of two. The new lower and upper uncertainties on $\sigma_{8}$ obtained are $0.15$ and $0.20$, respectively.

We then repeated the same procedure by increasing the covered sky area to $10^{4}\;\textrm{deg}^{2}$, as at the reach of Stage-IV surveys \citep{Euclid2011,Ivezi2019}, and increasing $N_{\textrm{obs}}$ by a proportional amount. This yielded lower and upper uncertainties for $\sigma_{8}$ equal to $0.11$ and $0.15$, respectively. This result shows a promising constraining power of EVS when applied to next-gen wide-field surveys, such as those proposed by ESA with \textit{Euclid} \citep{Euclid2022}.

Another important element to consider is the uncertainty on the observed cluster mass, which enters the likelihood both directly, through the first factor in equation \eqref{eq: EVS_likelihood}, and indirectly through the scale of the selection function. To determine the effects of reducing the observed cluster mass uncertainty, we repeated the analysis by considering an uncertainty of $\sim 20$ per cent for Abell 776's mass. We obtained a new uncertainty on $\sigma_{8}$ equal to $0.13$, which is comparable to the uncertainty obtained by increasing the covered sky area to $10^{4}\;\textrm{deg}^{2}$. This shows that a reduction of the observed mass uncertainty is another possible way to enhance the constraining power of EVS.

\section{Systematic errors}\label{sec: Systematic_errors}
Some major sources of systematic uncertainties are discussed in the following.

\subsection{Halo mass function model}\label{subsec: HMF_systematics}

A major source of systematic uncertainty is the modeling of the halo mass function. This uncertainty, mostly due to the variation of the high-mass tail of the mass function, was estimated by repeating the MCMC analysis with three different mass function definitions other than the \cite{Despali2016} mass function.

\begin{figure*}
\centering
\includegraphics[width = \textwidth]{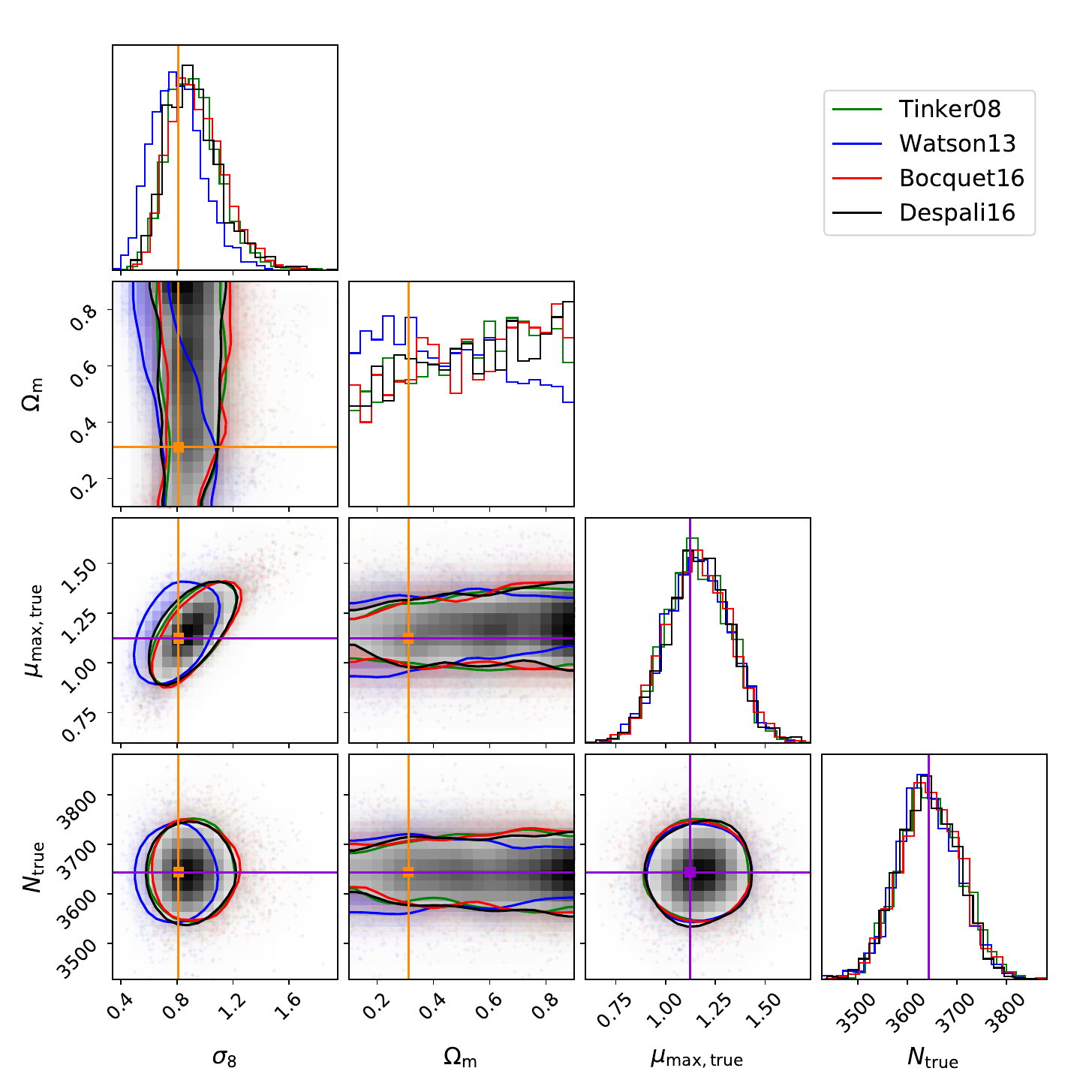}
\caption{\label{fig:corner_plot_hmf_uncertainty}Same as Fig. \ref{fig:corner_plot_fixed_ns}, but showing also the results obtained using different halo mass functions: \protect\cite{Tinker2008} (green), \protect\cite{Watson2013} (blue), \protect\cite{Bocquet2016} (red) and \protect\cite{Despali2016} (black). The \protect\cite{Planck18} results and the observed values $\mu_{\textrm{max, obs}}$ and $N_{\textrm{obs}}$ are shown in orange and violet, respectively.}
\end{figure*}

The mass functions considered were taken from \cite{Tinker2008}, \cite{Watson2013} and \cite{Bocquet2016}. The corner plot showing the comparison between the four results is shown in Fig. \ref{fig:corner_plot_hmf_uncertainty}, while the corresponding cosmological parameter estimates are reported in Table \ref{table: HMF_uncertainty_results}.

The difference between the highest and the lowest obtained central values for the parameters $\sigma_{8}$, $\Omega_{\textrm{m}}$ and $S_{8}$ is $(\Delta\sigma_{8})_{\textrm{HMF}} = 0.13$, $(\Delta \Omega_{\textrm{m}})_{\textrm{HMF}} = 0.10$ and $(\Delta S_{8})_{\textrm{HMF}} = 0.27$. These values can be taken as estimates of the systematic error due to the uncertainty on the halo mass function model assumption.

\begin{table}
\centering
\renewcommand*{\arraystretch}{1.50}

\caption{Comparison between the cosmological parameter estimated with the halo mass function described in \protect\cite{Tinker2008}, \protect\cite{Watson2013}, \protect\cite{Bocquet2016} and \protect\cite{Despali2016}, respectively.}

\begin{tabular}{cccc}
\hline
\hline
HMF definition & $\sigma_{8}$  & $\Omega_{\textrm{m}}$  & $S_{8}$ \\
\hline
Tinker-2008 & $0.91_{-0.17}^{+0.19}$ & $0.55_{-0.29}^{+0.23}$ & $1.19_{-0.36}^{+0.37}$\\
Watson-2013 &  $0.79_{-0.17}^{+0.19}$ & $0.46_{-0.24}^{+0.28}$ & $0.95_{-0.23}^{+0.22}$\\
Bocquet-2016 & $0.92_{-0.17}^{+0.21}$ & $0.56_{-0.29}^{+0.24}$ & $1.22_{-0.40}^{+0.43}$\\
Despali-2016 &$0.90_{-0.18}^{+0.20}$ & $0.54_{-0.27}^{+0.26}$ & $1.16_{-0.36}^{+0.40}$\\
\hline 
\end{tabular}

\label{table: HMF_uncertainty_results}
\end{table}

\subsection{Mass ranking}
A potential source of uncertainty is the misidentification of the most massive cluster. We considered Abell 776 as the most massive cluster of the catalog. Due to scatter and observational uncertainties, this might not be the case.

According to the WL mass calibration we adopted, the second most massive cluster of the catalog, J140101.92+025218, is located at redshift $z=0.25$ and has a mass equal to $M_{\textrm{max, obs, 2}} = (11.8\pm 4.4)\times 10^{14}h^{-1}\, \textrm{M}_{\odot}$, which is comparable to the observed mass value of Abell 776, $M_{\textrm{max, obs}} =(13.3\pm4.9)\times 10^{14}h^{-1}\,\textrm{M}_{\odot}$, and could thus be a candidate as the true most massive galaxy cluster of the sample. There could then be a systematic error contribution due to having underestimated the true value of the biggest mass measurement of the catalog.

To evaluate this contribution, we extracted $N=1000$ random samples from two Gaussian distributions, $\mathcal{G}(\mu_{\textrm{max, obs}}, \sigma_{\mu_{\textrm{max, obs}}})$ and $\mathcal{G}(\mu_{\textrm{max, obs, 2}}, \sigma_{\mu_{\textrm{max, obs, 2}}})$, associated to the first and second observed most massive cluster of the sample, respectively, and having a standard deviation given by the corresponding mass measurement uncertainty. We then constructed a new distribution by comparing term-wise the extracted samples and choosing the maximum between the two values. This corresponds to the distribution of $\textrm{max}[\mathcal{G}(\mu_{\textrm{max, obs}}, \sigma_{\mu_{\textrm{max, obs}}}),\, \mathcal{G}(\mu_{\textrm{max, obs, 2}}, \sigma_{\mu_{\textrm{max, obs, 2}}})]$.

\begin{figure}
\centering
\includegraphics[width=\columnwidth]{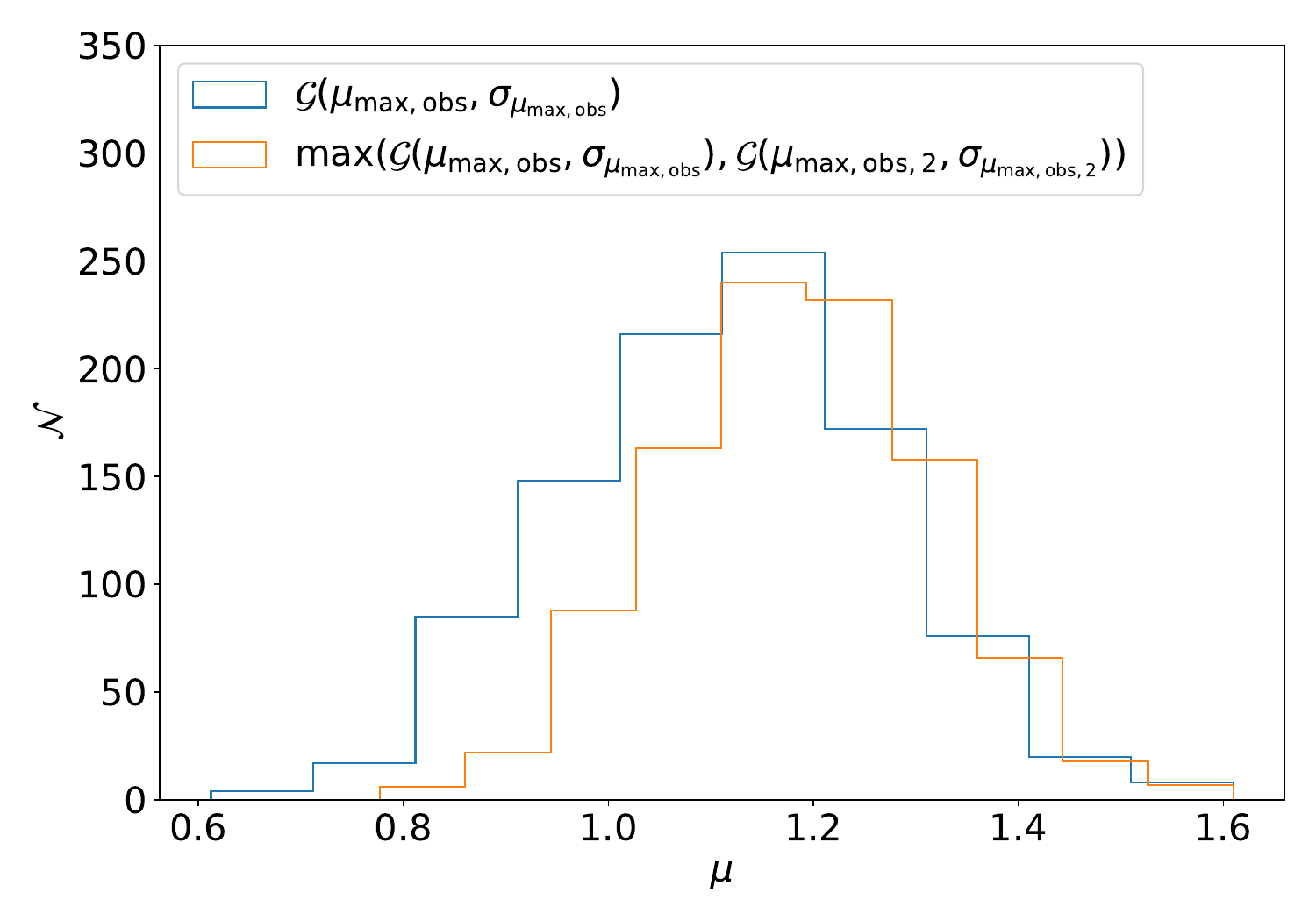}
\caption{\label{fig:systematics_mass_order}Comparison between the distribution associated to Abell 776's mass measurement, $\mathcal{G}(\mu_{\textrm{max, obs}}, \sigma_{\mu_{\textrm{max, obs}}})$, and the distribution $\textrm{max}[\mathcal{G}(\mu_{\textrm{max, obs}}, \sigma_{\mu_{\textrm{max, obs}}}),\, \mathcal{G}(\mu_{\textrm{max, obs, 2}}, \sigma_{\mu_{\textrm{max, obs, 2}}})]$.}
\end{figure}

We compared the mean of this new distribution to Abell 776's mass, obtaining a difference between the two of $\delta M = 0.07$, which is lower than the statistical uncertainty on the cluster masses. The comparison between the two distributions is shown in Fig. \ref{fig:systematics_mass_order}. We can consider this systematic error to have a negligible effect on the results.

\subsection{Reference cosmological parameters}\label{sec: Posterior_sampling_variable_ns}

\begin{figure*}
\centering
\includegraphics[width = \textwidth]{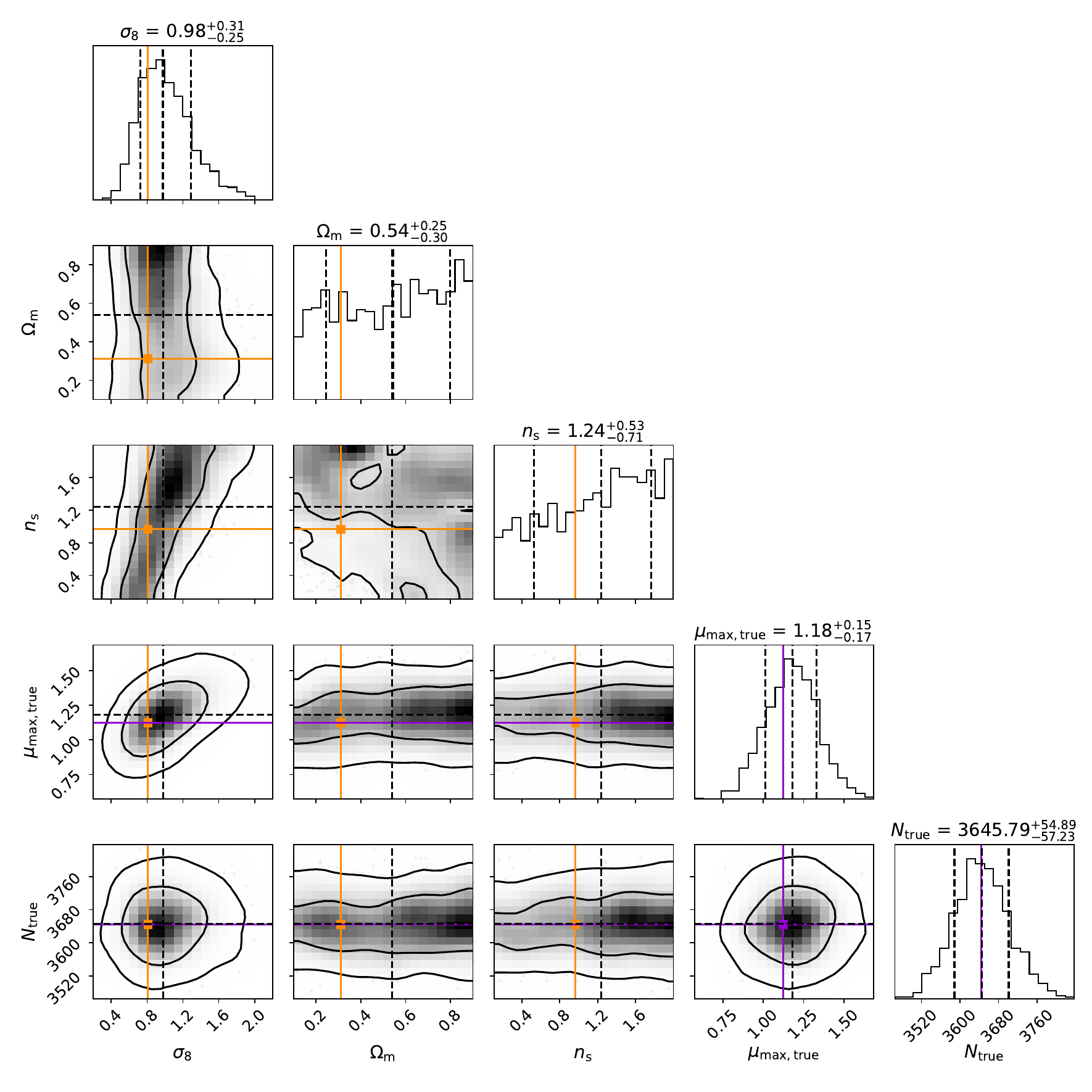}
\caption{\label{fig:corner_plot_ns_variable}Corner plot showing the 16th, 50th and 84th percentiles of the 1D parameter distributions with the $68.3$ per cent and $95.4$ per cent confidence regions for the corresponding 2D histograms, obtained by sampling the posterior probability distribution, considering the spectral index $n_{\textrm{s}}$ as a free parameter. Our results are shown with dashed black lines. The \protect\cite{Planck18} results and the observed values $\mu_{\textrm{max, obs}}$ and $N_{\textrm{obs}}$ are shown in orange and violet, respectively.}
\end{figure*}

In the procedure detailed in Section \ref{sec:Statistical_model}, we considered all the cosmological parameters other than $\Omega_{\textrm{m}}$ and $\sigma_{8}$ fixed to the values from \cite{Planck18}. To determine how much the results are robust with respect to a variation in these parameters, we repeated the analysis by letting one more parameter, the spectral index, vary. The results are shown in Fig. \ref{fig:corner_plot_ns_variable}, and the corresponding parameter estimates are reported in Table \ref{table: Results_variable_ns}.

A degeneracy in the $n_{\textrm{s}}$-$\sigma_{8}$ plane can be seen. Indeed, the Spearman linear correlation index between $n_{\textrm{s}}$ and $\sigma_{8}$, defined as the ratio between the covariance of the ranks associated to $n_{\textrm{s}}$ and $\sigma_{8}$ samples, $\textrm{cov}(R(n_{\textrm{s}}), R(\sigma_{8}))$, and the product of the respective rank uncertainties, $\sigma_{R(n_{\textrm{s}})}\cdot \sigma_{R(\sigma_{8})}$, is $\rho_{\textrm{s}} = 0.61$, with an associated p-value of $\gtrsim0$.

The results shown in Table \ref{table: Results_variable_ns} are consistent with the \cite{Planck18} results. The 84th percentile associated to $\sigma_{8}$ increased by $55$ per cent with respect to the value in Table \ref{table: Results_fixed_ns}, while the 16th percentile increased by $39$ per cent. The spectral index $n_{\textrm{s}}$ could not be significantly constrained.

\begin{table}
\centering
\renewcommand*{\arraystretch}{1.50}

\caption{Parameters associated to the likelihood distribution described in Section \ref{sec:Statistical_model}, estimated by considering the spectral index as a free parameter.}

\begin{tabular}{ccc}
\hline
\hline
Parameter & Prior & Posterior\\
\hline
$\sigma_{8}$ & $\mathcal{U}(0,2)$ & $0.98_{-0.25}^{+0.31}$\\
$\Omega_{\textrm{m}}$ & $\mathcal{U}(0.10,0.90)$ & $0.54_{-0.30}^{+0.25}$\\
$S_{8}$& $-$ & $1.26_{-0.41}^{+0.47}$\\
$n_{\textrm{s}}$ & $\mathcal{U}(0.1,2)$ & $1.24_{-0.71}^{+0.53}$\\
$\mu_{\textrm{max,true}}$&$\mathcal{U}(-2,4)$ & $1.18_{-0.17}^{+0.15}$\\
$N_{\textrm{true}}$ & $\mathcal{U}(0,10^{4})$ & $3646_{-57}^{+55}$\\
\hline 
\end{tabular}

\label{table: Results_variable_ns}
\end{table}

\section{Conclusions}\label{sec: Conclusions}
In this work, we estimated $\sigma_{8}$ using the extreme value statistics applied to galaxy clusters \citep{Waizmann2013}, by considering the probability of observing the most massive cluster of the AMICO KiDS-DR3 catalog \citep{Maturi2019}. By excluding all clusters outside the redshift interval $z\in[0.10, 0.60]$ and with an intrinsic richness $\lambda^{*} \leq 20$, the sample consists of $N_{\textrm{obs}} = 3644$  clusters, observed in an effective region of the sky equal to 377 $\textrm{deg}^{2}$. Our analysis accounted for the uncertainties associated to the largest cluster mass measurement and total cluster count measurement, respectively.

We obtained  $\sigma_{8}=0.90_{-0.18}^{+0.20}$. The result is compatible with the constraints obtained by \cite{Planck18} and previous AMICO KiDS-DR3 results \citep{Lesci2022, Lesci2022_clustering,Ingoglia2022} within $1\sigma$. Constraints on $\Omega_{\textrm{m}}$ are not significant. This is caused by the fact that the mass EVS is most sensitive to $\sigma_{8}$, rather than $\Omega_{\textrm{m}}$.

We also determined $S_{8} = 1.16_{-0.36}^{+0.40}$. This result, while being quite prior-dependent, is compatible within $1\sigma$ with measurements from \cite{Planck18}, \cite{Lesci2022} and \cite{Lesci2022_clustering}, and within $2\sigma$ with results from \cite{Ingoglia2022}.
    
The main source of systematic error comes from the uncertainty on the halo mass function model: by repeating the analysis with three other halo mass functions, \citep{Tinker2008, Watson2013, Bocquet2016}, we obtained an estimate of the systematic uncertainty by considering the difference between the highest and lowest central value for each parameter of interest. These uncertainties are equal to $(\Delta\sigma)_{\textrm{HMF}} = 0.13$, $(\Delta \Omega_{\textrm{m}})_{\textrm{HMF}} = 0.10$ and $(\Delta S_{8})_{\textrm{HMF}} = 0.27$, which are much lower than the respective statistical uncertainty contributions. The systematic uncertainties due to the uncertainty on the cluster mass ordering and due to having fixed all the other cosmological parameters are, instead, all negligible.

EVS of cluster masses is highly complementary to number count analyses. It is mostly sensitive to $\sigma_{8}$ rather than $S_{8}$, and joint analyses could break degeneracies between $\sigma_{8}$ and $\Omega_{\textrm{m}}$. Differently from number count analyses, selection function effects, purity and completeness play a small role in EVS. On the other hand, EVS is strongly affected by the theoretical uncertainty on the exponential tail of the halo mass function.

Our results have been obtained on a relatively small sample of galaxy clusters, located in a relatively small fraction of the sky. To test the effects of a larger catalog, we doubled the number of observed clusters, obtaining lower and upper uncertainties on $\sigma_{8}$ of $0.15$ and $0.20$, respectively. By increasing the covered sky area to $10^{4}\;\textrm{deg}^{2}$, as at the reach of Stage-IV surveys \citep{Euclid2011,Ivezi2019}, we instead obtained lower and upper uncertainties on $\sigma_{8}$ of $0.11$ and $0.15$.

Another route to obtain tighter constraints is to reduce the uncertainty on the cluster mass estimates. By halving the uncertainty on the mass estimate (i.e. considering a mass uncertainty of $\sim 20$ per cent), we indeed obtain an uncertainty on $\sigma_{8}$ of $0.13$, comparable to the reduction obtained due to increasing the covered sky area by 27 times its original value.

We plan to extend this analysis to the latest KiDS data release (DR5, Wright et al., in preparation), as well as DR4 \citep{Kuijken2019}, which covers an area of the sky of $\sim 10^{3}\; \textrm{deg}^{2}$ (almost 2.5 times larger than the one associated to the DR3), and \textit{Euclid} Data Release 1. An even more powerful application of our technique will indeed be possible with observations from the \textit{Euclid} Space Telescope \citep{Euclid2011} or the Vera C. Rubin Observatory \citep{Ivezi2019}, which are expected to detect thousands of distant, massive clusters (see e.g. \citealt{Sartoris2016}). Future surveys will also detect high mass clusters at very large redshifts ($z\gtrsim 1$), where the leverage of EVS can be very effective. Future works on this topic will also focus on understanding the potential of considering more than one cluster mass simultaneously to obtain the parameter constraints, as well as checking if there is any upper limit on the number of simultaneous clusters that one can consider for the joint constraints. We will also check under which conditions specifically the EVS approach is more competitive than other methods, e.g. cluster count or clustering.

\section*{Acknowledgments}
Based on data products from observations made with ESO Telescopes at the La Silla Paranal Observatory under programme IDs 177.A-3016, 177.A3017 and 177.A-3018, and on data products produced by Target/OmegaCEN, INAF-OACN, INAF-OAPD and the KiDS production team, on behalf of the KiDS consortium. MS acknowledges financial contributions from contract ASI-INAF n.2017-14-H.0 and contract INAF mainstream project 1.05.01.86.10. LM acknowledges support from the grants PRIN-MIUR 2017 WSCC32 and ASI n.2018-23-HH.0. All the cosmological functions used for the statistical analysis are built using the Python cosmology library \textsc{colossus}\footnote{\url{https://bdiemer.bitbucket.io/colossus/index.html}} \citep{Diemer2018}, while the MCMC analysis has been performed with the Python library \textsc{emcee}\footnote{\url{https://emcee.readthedocs.io/en/stable/}} \citep{ForemanMackey2013} and the resulting corner plots have been drawn using the Python library \textsc{corner.py} \footnote{\url{https://corner.readthedocs.io/en/stable/}} \citep{ForemanMackey2016}. We thank Konrad Kuijken for constructive criticism of the paper. We thank the anonymous referee for his comments, which helped us to better present our results.

\section*{Data Availability}
The data underlying this article will be shared on reasonable request to the corresponding author. The AMICO KiDS-DR3 cluster catalog is publicly available and can be accessed via the VizieR Online Data Catalog \citep{AKDR3Catalog2022}.


\bibliographystyle{mnras}
\bibliography{references}


\bsp 
\label{lastpage} 
\end{document}